**Distinguishing niche and neutral processes: issues in variation partitioning statistical methods and further perspectives**

Youhua Chen

Department of Renewable Resources, University of Alberta, Edmonton, T6G 2H1, Canada

Email: haydi@126.com or youhua@ualberta.ca

**Abstract**

Variance partitioning methods, which are built upon multivariate statistics, have been widely applied in different taxa and habitats in community ecology. Here, I performed a literature review on the development and application of the methods, and then discussed the limitation of available methods and the difficulties involved in sampling schemes. The central goal of the work is then to propose some potential practical methods that might help to overcome different issues of traditional least-square-based regression modeling. A variety of regression models has been considered for comparison. In initial simulations, I identified that generalized additive model (GAM) has the highest accuracy to predict variation components. Therefore, I argued that other advanced regression techniques, including the GAM and related models, could be utilized in variation partitioning for better quantifying the aggregation scenarios of species distribution.

**Introduction**

Variation decomposition in community ecology

It is quite often not only one process regulating and determining community structure.

Typically, the combination of multiple processes and their interactions will have profound impacts on resultant community structure. So, it is natural to ask which kinds of processes are dominant, while others are auxiliary. Thus, the variance in response variables can be separated into several parts, and by employing statistical methods, we can identify the contribution and relative importance of different ecological mechanisms.

Fig. 1 depicts the methods for performing variation decomposition at different data levels. The methods range from simple linear regression, to multiple regression models, to multivariate regression model, and other multivariate statistical methods.

How is variation partitioning related to the debate between niche and neutral processes?

Two mechanisms significantly affect species diversity patterns: neutral and deterministic factors. Environmental descriptors, defining the niche of species, are deterministic; while spatial descriptors, defining the dispersal ability of species, are neutral.

Since Hubbell's neutral theory (2001), a great amount of works tried to predict the power of neutral theory in empirical data. However, most of them failed to support neutral theory (McGill et al., 2006). There are several ways to test neutral theory. One is to generate individual predictions and test them by regression based on neutral theory. For example, the distance decay of species composition (Gilbert and Lechowicz, 2004); the priority effect of juvenile co-occurrence reduction (Gilbert et al., 2008).

Since the repellence of pure neutral theory in empirical test, the mainstream nowadays is to combine both niche and neutrality to explain community structure, and test the relative importance of each component. Therefore, variation partitioning is introduced (Borcard et al., 1992) to detect the contribution from each part of variations. In a work of Tuomisto and Ruokolainen (2006), they suggested that dispersal limitation derived from neutral process can only be tested by using distance-based methods.

How to perform variation partitioning?

To character species composition and environmental variables is a major topic in current ecological research. Redundancy analysis (suited for linear relationships between species composition and environmental variables) and Canonical correspondence analysis (handling nonlinear species-environment relationship) are the two widely used methods to investigate the relationship of environmental variables and species diversity information (Fig. 1). Variation partitioning can be used to test and determine the possibilities of individual predictors in influencing species distribution and abundance (Peres-Neto et al., 2006).

Variation partitioning can be divided into four parts: pure environmental variation, pure space variation, mixed environmental and space variation, and unexplained variation (Borcard et al., 1992). Figure 2 showed the schematic map of the variation components which were often complied in previous literature.

In principle, I use the spatial coordinates as the basic spatial descriptors. I can use the eigenvectors derived from the principal components of spatial coordinates, which has been used in some previous works (Dray et al., 2006). Or, I can get the Moran's eigenvector maps (Dray et al., 2006; Sattler et al., 2010), which is a general form of principal coordinates of neighbour matrices (Borcard and Legendre, 2002). Both methods use the eigenfunctions of spatial connectivity matrices, thus they are scale-independent.

When setting spatial descriptors as covariables, I can know the proportion of pure environmental variation. In contrast, when setting environmental variables as covariables, I can deduce the proportion of pure space variation. The mixed environmental and spatial variation can be derived from the subtraction of total known variation-pure environmental variation-pure spatial variation.

It is quite simple to perform comparative studies in ecological data by implementing variation partitioning in regression results. Basically, all kinds of statistical software and tools can implement variation partitioning, as long as they can perform multiple regression analysis. As I stated above, it needs only three times of running regressions, each of which should comprise environmental variables only, spatial variables only and both spatial and environmental variables together. From a convenience perspective, there is a commercial statistical software called Canoco (ter Braak and Smilauer, 2002), which is designed for constrained community ordination analysis. Variation partitioning can be implemented using partial CCA method in the package. Besides that, I can perform variation partitioning by using some open-source packages in R software. For example, the command "varpart" can be recalled to perform variation partitioning using "vegan" package (Oksanen et al., 2008). Other similar packages are also available, like 'varcan' (Pero-Neto et al., 2006).

Literature searching and study classification

To search the available publications relevant to separate the effects of niche processes and neutrality, I used the following databases: Google Scholar, Web of Science, Springlink,

Wiley-Blackwell, and Elsevier publishers.

The keywords used for query include "variation partition", "redundancy analysis", "niche and neutral processes", "dispersal limitation and habitat filtering", "spatial and environmental descriptors".

Based on the query results, I summarized two basic categories for the subject, which are 1) theoretical and methodological development; and 2) applications on different taxa and habitats. Table 1 summarized all the relevant work on the subject.

From the table, it seemed that most of applications of variation partitioning are on terrestrial ecosystems, and the studied taxa varied from birds to plants. Interestingly, there are only a few relevant works on marine and aquatic ecosystems, with only focus on fish assemblages.

A review of current methods and possible problems

There are many possible methods by applying nonlinear regression techniques to reveal the correlation of environmental and spatial variables against species distribution-composition matrix. The possibilities of introducing advanced nonlinear regression model, including local regression methods, general additive models and least partial square methods, can be beneficial to overcome the challenges.

Gilbert and Bennett (2010) performed a simulation comparison for analyzing the powers and differences among a variety of variation partitioning methods, most of which are widely used. Typically, the most prevailing methods are the regression on distance matrices (e.g., Mantel test), canonical correspondence analysis and redundancy analysis. Moreover, it is suggested to better retain spatial information by using some kinds of transformation called principal components of neighbour matrices (or Moran's eigenvector maps).

Despite their wide applications, Gilbert and Bennett (2010) found out that all kinds of tools have greatly underestimated each part of variances. For example, they found out that canonical ordination under-fitted the environmental variation, which was simulated in a high amount.

Potential methods

As found, the under-fit problem of different variation components by traditional ordination methods is largely due to the disability of traditional regression models. This is because all the available methods are built on the basis of least-square estimation of regression coefficients. All the conventional methods have the implicit assumption of homoscedasticity involved in the dataset more or less. Thus, as long as the data were composed of inherent heteroscedasticity, the power of least-square regression was questioned. Simple linear or nonlinear (e.g., polynomial regression and general linear models) fitting will be not possible to remove the impact of shifting data variance in the data set.

Fig. 3 showed the impacts of heteroscedasticity are hard to remove when plotting fitted residuals after conventional regression models. Thus, it sounds that a promising method to overcome the under-fit problem identified by Gilbert and Bennett (2010) is to adopt advanced regression models. Hence, in the following section of our synthesis is to propose advanced regression tools.

General additive model (GAM) and relevant nonlinear smoothing methods

The regression models in this category are of course nonlinear, however, another important feature is that they employed completely different ways aiming to solve the problem of heteroscedasticity. General additive model (GAM) typically has the power to remove the problem of variance heterogeneity, with the cost of difficult ecological explanation.

Here I generated a simple relationship between species abundance and one environmental variable with increasing variance across the landscape. Then, I applied different regression models to fit this heterogeneous variance case. The resultant residuals after fitting were as showed in Fig. 3, the simple linear model have only $R^2$-*adjusted*=0.4171, in contrast, GAM returned an $R^2$-*adjusted*= 0.542. Polynomial regression model is no more than a simple linear model, with $R^2$-*adjusted*=0.417. Moreover, when checking the regression residuals, I can find out that resultant residuals have no heterogeneity. In contrast, residuals from linear models (the same applied to quadratic nonlinear models, but not showed here) still have variance heterogeneity, and the case becomes worse at both lateral sides of the points.

**Implications and further perspectives**

Spatial heterogeneity and distributional aggregation may reduce the power of

Typically geographic coordinates are our only choices to measure spatial patterns and drivers of the community structure. However, sometimes it is hard to extract enough spatial information from simple geographic coordinates. In the case of multi-dimensional folding and transformation, Euclidean distances of geographic locations might not be sufficient to capture the variation caused by spatial distances. For example, species distributions typically show the aggregated, rather than random, patterns across different taxa. The driving reasons are usually dispersal-limited colonization and the constraint of habitat heterogeneity, and also biotic interactions, e.g., inter- and intra- specific competition (Cui et al., 2012).

Moran's spatial scales and edge effects

Sampling of different variables at different locations and scales might typically encounter the scale problem. The inconsistence of scales for different variables may lead to bias prediction on disentangling niche and neutrality processes. In such a case, the relative contribution of neutrality and niche processes driving the community structure may be misleading. Fig. 4 (upper graph) illustrated the scale issue when doing sampling in fields. Insufficient sampling across the region can give us a rough estimation of spatial gradient, but which is largely departed from the true gradient caused by middle-degree Moran's process. In this case, spatial variation should be overestimated. This issue can happen of course for environmental variables as well.

Edge effects may also inflate the possible separation of environmental filtering and spatial limitation. As showed in the same Fig. 4 (lower graph), if the sampling effect is focused on the transitional boundary areas of an environmental variable, the resultant explanation can be that the signaling of environmental filtering is not strong. It is easy to avoid the edge problem for one environmental variable. However, for the case of multiple variables, as it should be hard to predict their transitional boundaries, the sampling scheme can be always coupled with edge effects.

At another side, as known that, both processes can have similar predictions on many facets of community structures. For example, niche process can generate the same distance-decaying pattern as that of neutrality process. In the case like that, typically the resultant community structure is co-dominated by both mechanisms and hard to separate without additional information about the community. Thus, it might be not effective to use partial regression techniques to separate niche and neutrality processes.

Variable selection process

Maybe variable selection is an improved way to better capture the most correlated variation information for spatial and environmental drivers. In tradition, backward or forward variable selection procedure is applied to choose the optimal subsets of variables. In the case of variation partitioning, I can do the variable selection separately, then obtain the most correlated spatial and environmental predictors to perform variation decomposition. This method did provide the most significant variables, but a small change of variable subset will lead to a great amount of changes in the resulting variation. Thus, the discrete process may reduce the prediction accuracy when new variables are included (Tibshirani 1996). Fortunately, I have other advanced model selection methods, for example, Lasso and ridge regressions can be good alternatives for choosing good candidates in variation partitioning.

An integrative way to partition and understand ecological communities

The explanation is the most challenging thing for general additive models, although it has a higher appealing prediction power than linear models. The comprising manner is to use simple linear or nonlinear regression to fit the data when obtaining the possible curve pattern inspired by GAM models. Then, combining the fitted adjusted $R^2$ from GAM and fitted line from either linear or simple nonlinear regression models, I can clearly and easily explain the possible relationships and mechanisms that might dominate the community data.

1010

Table 1 A literature review on the methodology and applications of variation partitioning in community ecology

| Sub-discipline | Description | Literature |
| --- | --- | --- |
| 1, Methodological background and development | Advocating or criticizing the variance partitioning and developing relevant tools | 1, the original paper describing variation partitioning: Borcard et al. (1992)<br>2, comparison of different methods on performing variation partitioning. For example, Mantel test, multiple regression model, canonical correspondence analysis and so on (Legendre et al. 2005, Legendre, 2008)<br>3, ecological questions that can be addressed by variation partitioning: beta diversity (Legendre et al., 2005; Legendre et al., 2009); neutrality versus niche (e.g., Smith and Lundholm, 2010; Tuomisto et al., 2012);<br>4, rebuttal to variation partitioning methods and relevant technical aspects:<br>Tuomisto and Ruokolainen (2006) suggested that distance method (like Mantel test) should be the only choice to test neutral hypothesis.<br>Gilbert and Bennett (2010) found that traditional variation partitioning methods have a very restricted power to correctly quantify each part of variances involved in the simulated data.<br>Diniz-Filho et al. (2012) suggested that spatial autocorrelation test (Moran's I index) can be linked to niche and neutrality partitioning. |
| 2, Applications on various taxonomic hierarchy and spatial scales | Applying variation partitioning to different community assemblages across different ecosystems, areas and taxa. | 1, terrestrial ecosystems:<br>Oribatid mites (Borcard and Legendre, 1994; Lindo and Winchester, 2004);<br>Forest birds (e.g., Cushman and McGarigal, 2002; Pearman, 2002);<br>Pteridophyte plants (Jones et al., 2008)<br>2, marine and aquatic ecosystems:<br>Stream fish community (Stward-Koster et al., 2007);<br>Pelagic fish assemblages (Peltonen et al., 2007) |

Figure 1 Different methods, ranging from a simple regression to multivariate statistics, have been used for variation partitioning to distinguish environmental and spatial filtering. Solid arrow denotes that the method can be used for performing variation partitioning tests, while dashed arrow denotes that that method (pointed by the head of arrow) can be generalized or deduced from another method.

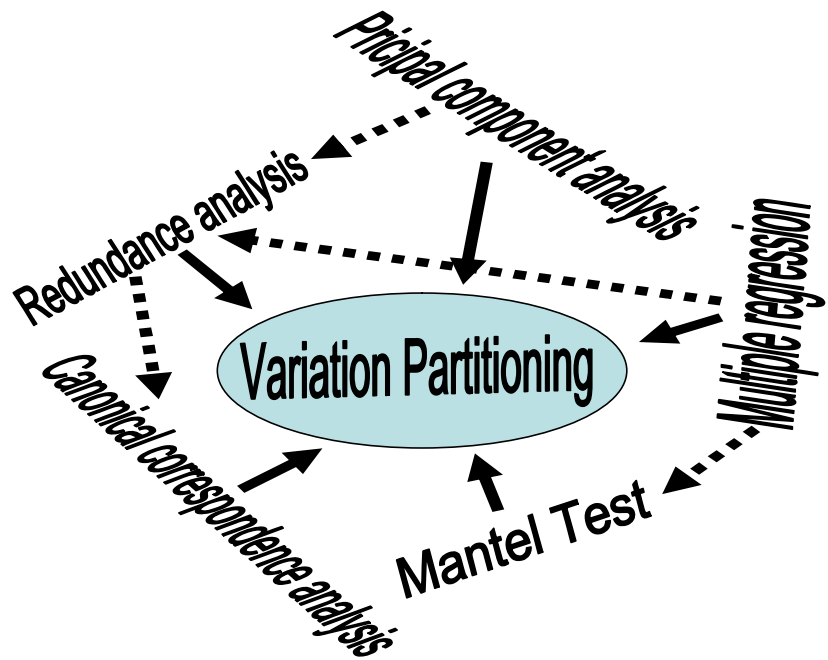

Figure 2 Schematic map showing different components and fractions that are related to variation partitioning. a-pure environmental variation; b-mixed environmental and space variation; c-pure space variation; d-unexplained variation. a+b+c+d=total variance involved in the community data.

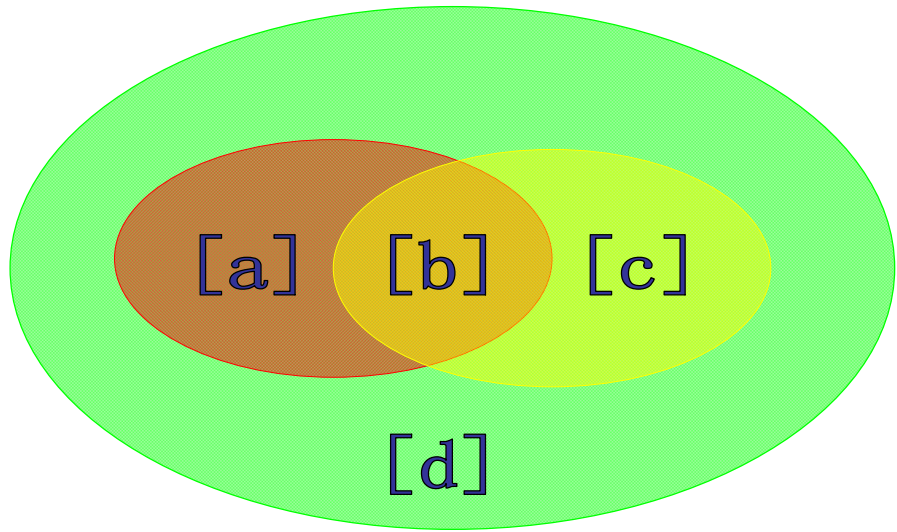

Figure 3 The problem of variance heterogeneity and the effectiveness of additive models compared to linear models to remove variance heterogeneity (left figure is the residuals fitted by simple linear model, while the residuals derived from general additive fitting with Gaussian family is showed in the right figure). It is clear a humped pattern (non-homogenous residuals) still occurs in the left graph.

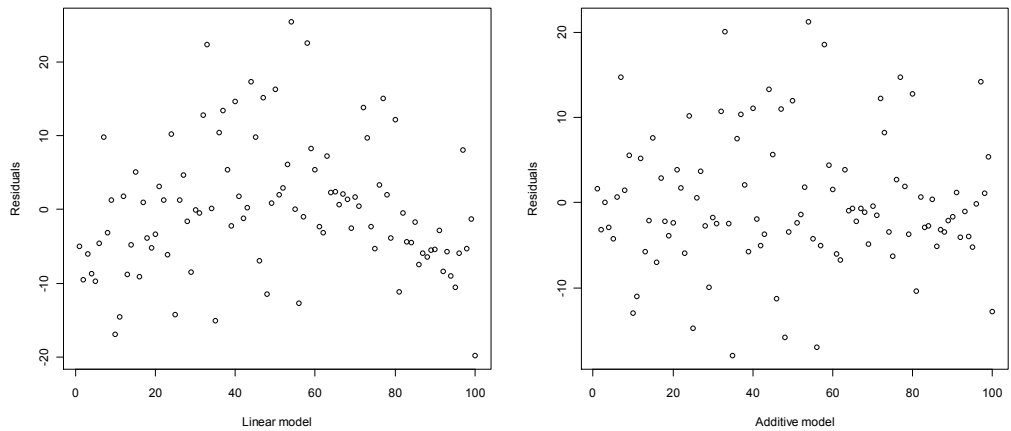

Figure 4 Sampling biases caused by Moran's spatial scales and edge effects. Two rectangular areas indicate two distinct levels of an environmental variable (e.g., precipitation, elevation, temperature and so on). Square and triangle points with red and blue colors indicated two species. Ellipse circles represent density of species population. Transparent gray squares represent sampling plots across the region. In principle, the community bounded by the large square is structured by environmental filtering. The upper graph illustrates the insufficient sampling case, which make the wrong conclusion that environmental filtering is not important to capture the beta-diversity. The lower graph illustrate the edge sampling case, which make the wrong conclusion that spatial filtering is much more important than environmental filtering to structure the community.

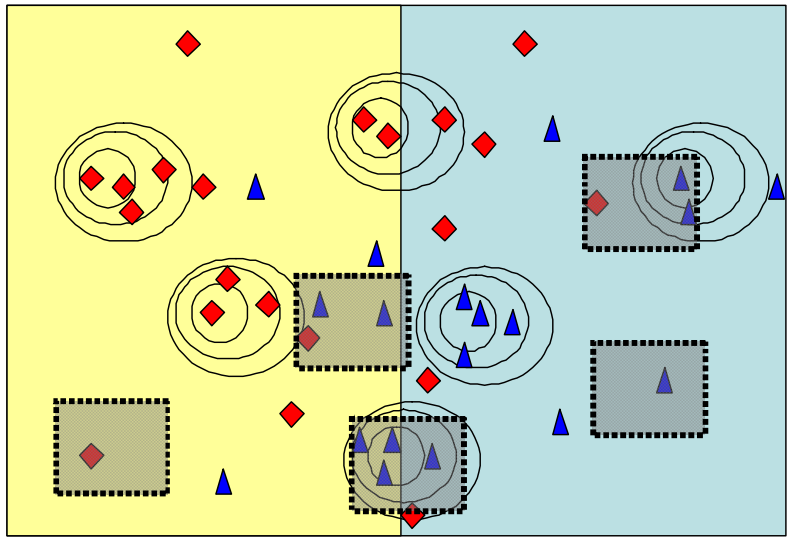

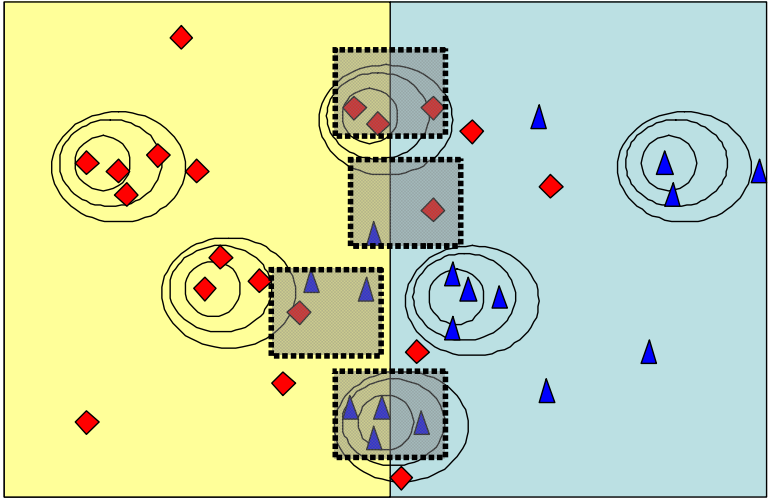